\def\be{\begin{equation}}
\def\ee{\end{equation}}
\def\bea{\begin{eqnarray}}
\def\eea{\end{eqnarray}}
\def\bml{\begin{mathletters}}
\def\eml{\end{mathletters}}
\def\b{\bibitem}
\documentstyle[aps,prl,epsf,floats]{revtex}

\begin{document}
\def\SNG{{\em Physical Review Style and Notation Guide}}
\def\LUG {{\em \LaTeX{} User's Guide \& Reference Manual}}
\def\btt#1{{\tt$\backslash$\string#1}}%
\def\REVTeX{REV\TeX}
\def\AmS{{\protect\the\textfont2
        A\kern-.1667em\lower.5ex\hbox{M}\kern-.125emS}}
\def\AmSLaTeX{\AmS-\LaTeX}
\def\BibTeX{\rm B{\sc ib}\TeX}
\twocolumn[\hsize\textwidth\columnwidth\hsize\csname@twocolumnfalse%
\endcsname
\centerline{\small Volume 87, Paper \# 127003\hskip 88pt PHYSICAL REVIEW 
                   LETTERS \hfill 2001}
\vskip 3pt
\hrule
\vskip 5pt
\title{Strong Enhancement of Superconducting $T_c$ in Ferromagnetic Phases} 
\author{T.R. Kirkpatrick$^1$, D. Belitz$^2$, Thomas Vojta$^{3,4}$, 
        and R. Narayanan$^3$}
\address{$^1$Institute for Physical Science and Technology, and Department of
              Physics, University of Maryland,\\ College Park, MD 20742\\
         $^2$Department of Physics, and Materials Science Institute,
         University of Oregon, Eugene, OR 97403\\
         $^3$Department of Physics, University of Oxford, 1 Keble Rd, Oxford OX1 3NP, UK\\
         $^4$Institut f{\"u}r Physik, TU Chemnitz, D-09107 Chemnitz, FRG}
\date{\today}
\maketitle
\begin{abstract}
It is shown that the critical temperature for spin-triplet, $p$-wave
superconductivity mediated by spin fluctuations is generically much
higher in a Heisenberg ferromagnetic phase than in a paramagnetic one,
due to the coupling of the magnons to the
longitudinal magnetic susceptibility. 
Together with the tendency of the low-temperature ferromagnetic
transition in very clean Heisenberg magnets to be of first order, this
qualitatively explains the phase diagram recently observed in UGe$_2$.
%
%
\end{abstract}
\pacs{PACS numbers: } 
]
It has long been known that, in principle, the exchange of magnetic 
fluctuations between electrons can induce 
superconductivity\cite{AndersonBrinkman}. Magnetic
fluctuations become large in the vicinity of continuous magnetic
phase transitions, which makes nearly ferromagnetic materials, or 
ferromagnets with a low Curie temperature, natural candidates for this
phenomenon. In contrast
to the much more common phonon-exchange case, which usually leads to
electron pairing of spin-singlet, s-wave nature, the magnetically mediated
pairing is strongest in the spin-triplet, p-wave channel. p-wave
superconductivity is very sensitive to nonmagnetic 
impurities and
therefore can be expected only in extremely pure samples. The combined
requirements of high purity, low temperatures, and vicinity to a
ferromagnetic transition severely restrict the number of promising materials.
Indeed, until recently
there were no convincing simple examples of magnetically induced
superconductivity, and the paramagnon interpretation of superfluid
$^3$He\cite{AndersonBrinkman,LevinValls}
was considered the best example of pairing by exchange of magnetic
fluctuations.

This situation has recently changed, due to the observation of the coexistence
of ferromagnetism and superconductivity in UGe$_2$\cite{Saxena_etal}. In
contrast to other uranium compounds, UGe$_2$ has more in common with classic
d-electron ferromagnets, like Fe, Co, and Ni, than with heavy-fermion systems. 
The persistence of ferromagnetic order within the superconducting phase has 
been ascertained by means of neutron scattering, and the itinerant 
ferromagnetism and the superconductivity are believed to arise from the same
electrons\cite{Saxena_etal}. Since superconductivity in the presence of
ferromagnetism must be of spin-triplet type, magnetically induced pairing is
an obvious candidate for the observed superconductivity,
although a phonon mechanism has also been proposed\cite{ShimaharaKohmoto}.

\begin{figure}[ht]
\epsfxsize=85mm
\centerline{\epsffile{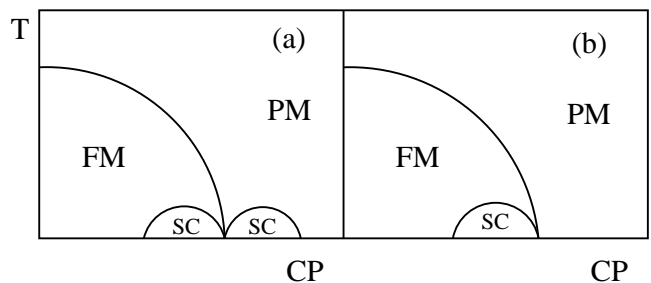}}
\vskip 5mm
\caption{Schematic phase diagram showing the paramagnetic (PM), ferromagnetic
 (FM), and superconducting phases (SC) in a temperature (T) - control parameter
 (CP) plane. (a) shows the qualitative prediction of paramagnon 
 theory\protect\cite{FayAppel}, and (b) qualitatively shows the phase diagram 
 as observed in UGe$_2$\protect\cite{Saxena_etal} and explained by the theory
 presented here. In 
 Ref.\ \protect\onlinecite{Saxena_etal}, hydrostatic pressure serves as CP.}
\vskip 0mm
\label{fig:1}
\end{figure}

The nature of the phase diagram reported in Ref.\ \onlinecite{Saxena_etal}
is, however, not obviously consistent with existing models of spin fluctuation
induced superconductivity, see Fig.\ \ref{fig:1}. 
Fay and Appel\cite{FayAppel} have calculated the superconducting
$T_c$ for a p-wave, equal-spin pairing state in both the 
paramagnetic (PM) and ferromagnetic (FM) phases close to a continuous magnetic
transition. Using a McMillan-type formula, they found values of $T_c$ 
on either side of the transition that are within 20\% of one another. Their 
shape of $T_c$ as a function of the distance $t$ from the magnetic 
transition is very similar to that obtained by Levin and 
Valls\cite{LevinValls}, who solved the Eliashberg equations numerically
in the PM phase, although the absolute values of $T_c$ are smaller
in the McMillan approximation. More recently,
Roussev and Millis\cite{RoussevMillis} have obtained similar results in the
PM phase. Contrary to this theoretical 
expectation of a superconducting phase diagram that is roughly symmetrical
with respect to the magnetic phase boundary, Fig.\ \ref{fig:1}(a), 
the authors of Ref.\ \onlinecite{Saxena_etal} observed superconductivity 
at temperatures up to about 500 mK within the FM phase 
only, Fig.\ \ref{fig:1}(b). 
Qualitatively the same phase diagram has very recently been observed in
ZrZn$_2$\cite{Pfleiderer_et_al}.
Since the spin fluctuations become large on either
side of the magnetic transition, it seems hard
to reconcile this experimental result with paramagnon 
theory\cite{semantics_footnote}.

In this Letter we show that the observed phase diagram can nevertheless 
be understood in these terms. The key 
lies in the existence of spin waves or magnons in the FM phase,
which couple to the longitudinal susceptibility and 
contribute a mode-mode coupling term to the latter that has no analog 
in the PM phase.
We will see that this produces a superconducting transition temperature
which under reasonable assumptions can easily be 50 times larger than in the
PM phase.

\begin{figure}[htb]
\epsfxsize=85mm 
\centerline{\epsffile{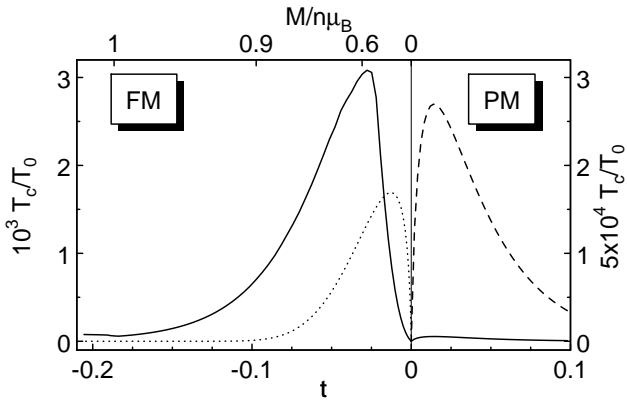}}
\vskip 1mm
\caption{Superconducting $T_c$ (solid curve, left scale) as a function 
 of the distance from the critical point $t$, and the magnetization $M$.
 The dashed line (right scale) shows $T_c$ in the PM phase scaled by a
 factor of 50, and the dotted curve (right scale) is the result in the FM
 phase without the mode-mode coupling effect. See the text for further 
 explanation.}
\vskip -1mm
\label{fig:2}
\end{figure}

We have included this effect in a McMillan-type $T_c$ calculation similar
to the one in Ref.\ \onlinecite{FayAppel} A representative result of our 
analysis is shown in Fig.\ \ref{fig:2}.
The solid curve (left-hand scale) represents the superconducting $T_c$ as a
function of the dimensionless distance $t$ from the FM critical 
point. Also shown is the magnetization $M$ in the FM phase, in units of the
saturation magnetization $\mu_B\,n$, with $n$ the electron number density
and $\mu_B$ the Bohr magneton. $T_c$ is measured in units of a characteristic
temperature $T_0$ that is given by either the Fermi temperature or a band
width, depending on the model considered. The dashed curve shows the result
in the PM phase scaled by a factor of 50 (right-hand scale), 
and the dotted curve in the FM phase (also scaled by a factor of 50,
right-hand scale) represents the result 
that is obtained upon neglecting the mode-mode
coupling effect. 
Notice that the maximum $T_c$ in the FM phase is
more than 50 times higher than in the PM phase. This
{\em relative} difference between $T_c$ in the two phases is the important
result of our analysis. The absolute values should not be taken very
seriously, as calculating $T_c$ is notoriously difficult and our simple
mean-field treatment is certainly not adequate for this purpose. However,
the relative comparison we expect to be reliable. It predicts a pronounced
asymmetry between the PM and FM phases, which in the case of UGe$_2$ means
that superconductivity in the PM phase should not be expected at
temperatures above at most 10 mK, in agreement with the experiment.
Also of interest is the fact that at low temperatures
the FM transition in very clean itinerant Heisenberg systems is
generically of first order, as has been predicted
theoretically\cite{us_1st_order} and is indeed observed in 
UGe$_2$\cite{Saxena_etal} as well
as in MnSi\cite{Pfleiderer_etal}. For the purpose of our discussion
this simply means that values of $|t|$ smaller than some minimum
value are not experimentally accessible.

In the remainder of this Letter we sketch the theoretical analysis that
has led to these results. For an order parameter (OP) field, we choose 
${\cal F}(x,y) = \psi_{\uparrow}(x)\,\psi_{\uparrow}(y)$, 
with $\psi_{\sigma}(x)$ an electronic field
with spin index $\sigma$ and space-time index $x$\cite{beta_phase_footnote}. 
The OP, i.e. the expectation value $\langle{\cal F}(x,y)\rangle = F(x-y)$,
is the anomalous Green function. The orbital symmetry of the OP is still
unspecified, we will later choose the p-wave case. 

We have derived coupled equations of motion for $F$ and the normal
Green function, $G$, that lead to a loop expansion for the equation
of state\cite{us_tpb}. Our model is a microscopic 
action $S$ with a free-electron part, $S_0$, and spin-singlet and 
spin-triplet interaction terms, 
\be
S_{\rm int}^t = \frac{\Gamma_t}{2} \int dx \bigl[{\bf n}_s(x)\bigr]^2
                ,\ 
S_{\rm int}^s = \frac{-\Gamma_s}{2} \int dx\bigl[n_c(x)\bigr]^2 .
\label{eq:1}
\ee
Here ${\bf n}_s(x)$ and $n_c(x)$ are the electronic spin and charge density
fields, respectively, and $\Gamma_t$ and $\Gamma_s$ are the spin-triplet
and spin-singlet interaction amplitudes. We assume that
screening has been built into the starting action, so the interaction
amplitudes are simply numbers. By putting
$\Gamma_s = \Gamma_t$ one obtains the Hubbard model considered in
Ref.\ \onlinecite{FayAppel}. The magnetic
equation of state we treat in zero-loop approximation.
The superconducting
equation of state needs to be calculated in one-loop approximation 
in order to capture the spin-fluctuation induced pairing. 
It takes the form of linearized
strong-coupling equations that are similar to those in 
Ref.\ \onlinecite{FayAppel}. These equations
can be rewritten as an eigenvalue problem, which can then be solved 
numerically, using some theory for the (para)magnon propagators as input. This 
is the established procedure to calculate the critical temperature for
phonon-mediated superconductivity\cite{AllenDynes}, and it
has been employed in the case of magnetically induced superconductivity
or superfluidity in Refs.\ \onlinecite{LevinValls,RoussevMillis}.

Even with a complete numerical solution of the strong-coupling equations,
the superconducting $T_c$ is notoriously hard to predict.
This holds {\it a fortiori} in the case of magnetically mediated
superconductivity since (1) there is much 
less experimental information about the paramagnon propagator, that could 
be used as input, than about
phonon spectra, and (2) there is no
analog of Migdal's theorem. Our ambition here is therefore {\em not}
to calculate $T_c$, but rather to perform a {\em relative}
comparison of $T_c$ values in the PM and FM phases, respectively. 
For this purpose, a simple McMillan-type
approximation for $T_c$\cite{AllenDynes} suffices. We obtain
\be
T_c = T_0(t)\,\exp\left[-(1 + d_0^L + 2d_0^T)/d_1^L\right]\quad.
\label{eq:3}
\ee
Here $T_0(t)$ is a temperature scale that will be specified below.
Specializing to the p-wave case, the $d_{0,1}^{L,T}$ read
\bml
\label{eqs:4}
\bea
d_1^L&=&\frac{\Gamma_t N_F^{\uparrow}}{(k_F^{\uparrow})^2}
   \int_{0}^{2k_F^{\uparrow}} dk\,k\,\left(1 - \frac{k^2}{2(k_F^{\uparrow})^2}
      \right)\,D_L(k,i0)\ ,
\label{eq:4a} \\
d_0^L&=&\frac{\Gamma_t N_F^{\uparrow}}{(k_F^{\uparrow})^2}
   \int_{0}^{2k_F^{\uparrow}} dk\,k\,D_L(k,i0)\quad,
\label{eq:4b}\\
d_0^T&=&\frac{\Gamma_t N_F^{\uparrow}}{(k_F^{\uparrow})^2}
   \int_{k_F^{\uparrow}-k_F^{\downarrow}}^{k_F^{\uparrow}+k_F^{\downarrow}}
      dk\,k\,D_T(k,i0)\quad.
\label{eq:4c}
\eea
\eml%
$k_F^{\uparrow} (k_F^{\downarrow})$ are the Fermi wavenumbers for the
up (down)-spin Fermi surface, and $N_F^{\uparrow}$ is the density of states
at the up-spin Fermi surface. In the PM phase,
$k_F^{\uparrow} = k_F^{\downarrow} \equiv k_F$. $D_{L,T}(q)$ are the
longitudinal and transverse (para)magnon propagators. They are related to
the electronic spin susceptibility $\chi$ via
$D_{L,T}(q) = \chi_{L,T}(q)/2N_F$, with $N_F$ the density of states at
the Fermi level in the PM phase. We use the
expressions that were derived in Ref.\ \onlinecite{us_fm_mit}, with one
crucial modification that we will discuss below. From that paper we
obtain in the PM phase, and in the limit of small wavenumbers,
\be
D_{L,T}(q,i0) = 1/[t + b_{L,T}(q/2k_F)^2]\quad,
\label{eq:5}
\ee
In the Gaussian approximation of Ref.\ \onlinecite{us_fm_mit},
$b_L = b_T = 1/3$. However, there is no reason to
prefer this Gaussian approximation over any other approximation scheme. 
The functional form of the
long-wavelength expression, Eq.\ (\ref{eq:5}), on the other hand,
is generic. We therefore adopt Eq. (\ref{eq:5}) with  $b_{L,T}$ 
arbitrary coefficients of $O(1)$. By the same reasoning, we have in
the FM phase, in the limit of long wavelengths and small 
frequencies,
\bml
\label{eqs:6}
\bea
D_L(q,i0)&=&1/[5\vert t\vert/4 + b_L (q/2k_F)^2]\quad,
\label{eq:6a}\\
D_T(q,i\Omega)&=&\frac{\Delta/4\epsilon_F}{(1-t)^2}\,
   \left(\frac{1}{i\Omega/4\epsilon_F + (\Delta/2\epsilon_F)b_T (q/2k_F)^2}
         \right.
\nonumber\\
&&\hskip 1pt - \left.\frac{1}{i\Omega/4\epsilon_F - (\Delta/2\epsilon_F)b_T 
     (q/2k_F)^2}\right) \quad,
\label{eq:6b}
\eea
\eml%
with $\Delta$ the Stoner band splitting. For $0<\Delta<n\Gamma_t$,
$\Delta$ is related to the magnetization $M$ by 
$M = \mu_{\rm B}\Delta/\Gamma_t$.

Two comments: (1) In a strict long-wavelength expansion
of the propagators from Ref.\ \onlinecite{us_fm_mit} the $b_L$ and $b_T$ 
in Eqs.\ (\ref{eq:6a},\ref{eq:6b}) become magnetization dependent.
We ignore this effect and use the same values as in the 
PM phase. We have compared this approximation against using the
full propagators from Ref.\ \onlinecite{us_fm_mit}, see below.
(2) The factor of 5/4 in Eq.\ (\ref{eq:6a}) arises since we
keep the particle number density fixed, as is the case experimentally,
rather than the chemical potential, see Ref.\ \onlinecite{FayAppel}.

\begin{figure}[htb]
\epsfxsize=40mm
\centerline{\epsffile{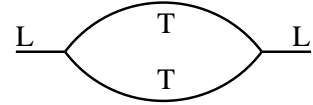}}
\vskip 5mm
\caption{Mode-mode coupling contribution to the longitudinal (L) propagator
 $D_L$ from the transverse (T) ones.}
 \vskip 0mm
\label{fig:3}
\end{figure}

We now consider the longitudinal magnetic propagator in the FM
phase in more detail. In a Heisenberg ferromagnet, the transverse spin 
waves or magnons couple to the longitudinal susceptibility $\chi_L$.  
This effect is most easily demonstrated within a nonlinear sigma-model
description of the ferromagnet, which treats the 
order parameter ${\bf M}$ as a vector of fixed length $M$, and 
parameterizes it as
${\bf M} = M\,(\sigma(x),\pi_1(x),\pi_2(x))$ with 
$\sigma^2 + \pi_1^2 + \pi_2^2 = 1$,
$M$ the magnetization, and $x$ a space-time index\cite{ZJ,BKMV}. 
The diagonal part of the $\pi_i$-$\pi_j$ propagator,
$\langle\pi_i\pi_i\rangle = (M^2/2N_F)\,D_T$,
is proportional to the transverse propagator $D_T$, and the off-diagonal
part has been calculated in Ref.\ \onlinecite{BKMV}.
The longitudinal propagator, $D_L = (M^2/2N_F)\langle\sigma(x)\sigma(y)\rangle$,
can be expanded in a series of $\pi$-correlation
functions,
\bea
\langle\sigma(x)\sigma(y)\rangle &=& 
 1 - 2\langle\pi_i(x)\pi^i(x)\rangle
\nonumber\\
&& \hskip 10pt + \langle\pi_i(x)\pi^i(x)\pi_j(y)\pi^j(y)\rangle + \ldots
\label{eq:6'}
\eea
where repeated indices are summed over.
At one-loop order, the term of order $\pi^4$ yields the diagram 
shown in Fig.\ \ref{fig:3}. Notice that the sigma model, which
neglects all longitudinal fluctuations, replaces the external legs
by constants.
Power counting shows that at nonzero temperature, and for dimensions
$d<4$, this contribution causes the homogeneous longitudinal susceptibility
to diverge everywhere in the FM phase\cite{BrezinWallace}. More generally,
this one-loop contribution, together with the zero-loop one, Eq.\ (\ref{eq:6a}),
yields a functional form for $D_L$ in the FM phase that is exact at small
wavenumbers. This diagram has no analog in the PM phase, while
all other renormalizations of the propagators will give comparable 
contributions in both the PM and FM phases.
It is therefore reasonable to calculate $T_c$ based on a one-loop approximation
in the FM phase, and compare it to a zero-loop calculation in the PM 
phase. We have used Eq.\ (\ref{eq:6b}) for the internal propagators in 
Fig.\ \ref{fig:3}. Since the coupling constants involve a
wavenumber integral, Eqs.\ (\ref{eqs:4}), we also need to go beyond the
sigma model and keep the wavenumber dependence of the external ones. For
computational simplicity, we have modeled the external legs by 
replacing Eq.\ (\ref{eq:6a}) with a step function that cuts off the
momentum integral at $k/2k_F = \sqrt{5\vert t\vert/4b_L}$. With these
approximations, the momentum integral in Fig.\ \ref{fig:3} can be
done analytically, leaving the frequency sum to be performed
numerically. The result, and the resulting contribution to $d_1^L$
and $d_0^L$, depend on the temperature, so Eq. (\ref{eq:3})
now needs to be solved self-consistently for $T_c$. 

We still need to specify the temperature scale $T_0(t)$.
Following Ref.\ \onlinecite{FayAppel}, we use the
prefactor of $\vert t\vert$ in Eqs.\ (\ref{eq:5},\ref{eq:6a})
as a rough measure of the magnetic excitation energy, 
\be
T_0(t) = T_0\,[\Theta(t)\,t + \Theta(-t)\,5\vert t\vert/4]\quad,
\label{eq:7}
\ee
with $T_0$ a microscopic temperature scale that is related to the Fermi
temperature (for free electrons) or a band width (for band electrons).
This qualitatively reflects the suppression of the
superconducting $T_c$ near the FM transition due to effective
mass effects\cite{LevinValls,FayAppel,RoussevMillis}. 

\begin{figure}[thb]
\epsfxsize=85mm
\centerline{\epsffile{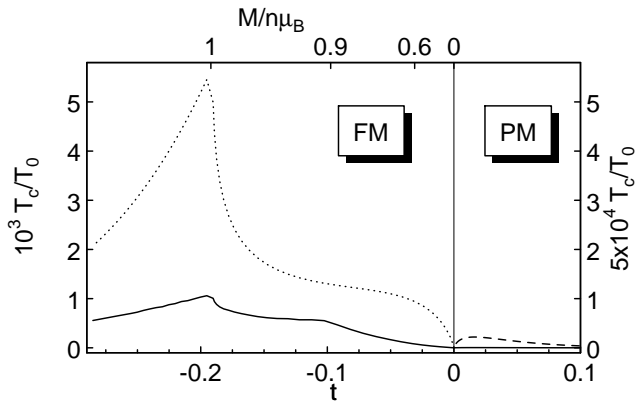}}
\vskip 1mm
\caption{Same as Fig.\ \ref{fig:2}, but for different parameter values
         (see the text).}
\vskip -2mm
\label{fig:4}
\end{figure}

We are now in a position to choose parameters and calculate explicit
results. We put $\Gamma_s = \Gamma_t$\cite{FayAppel};
other reasonable choices yield similar results.
Let us first ignore the mode-mode coupling contribution to $d_1^L$ and $d_0^L$.
We have performed the
calculation both with the full propagators from Ref.\ \onlinecite{us_fm_mit}
and with the schematic Landau propagators, Eqs.\ (\ref{eq:5}) and
(\ref{eq:6a}). With $b_L = 0.23$, $b_T = 0.4$ the two results
are within 10\% of one another, and also very similar to those obtained
by Fay and Appel\cite{FayAppel}. We then use these values of $b_{L,T}$
to calculate the mode-mode coupling contribution, and solve the 
$T_c$-equation. The result is shown in Fig.\ \ref{fig:1}
and has been discussed above. We have also explored the effect of varying 
the parameters $b_{L,T}$. With $b_L = b_T = 1$
we obtain the result shown in Fig.\ \ref{fig:4}. The
(unphysical) zero-loop result in the FM phase is very sensitive to the 
parameters, while the enhancement of the (physical) one-loop result over
the $T_c$ in the PM phase is rather robust. However, the position of the
maximum of $T_c$ changes compared to Fig.\ \ref{fig:1}; it now
occurs at the point where the 
magnetization reaches its saturation value. The reason is as
follows. As one approaches the magnetization saturation point from low
magnetization values, the transverse coupling constant $d_0^T$ vanishes,
and remains zero in the saturated region. Effectively, the Heisenberg
system turns into the Ising model discussed in 
Ref.\ \onlinecite{RoussevMillis}. If the longitudinal coupling
constant $d_1^L$ still has a substantial value at that point, then this
leads to an increase in $T_c$. This is a very strong effect in the zero-loop
contribution, see Fig.\ \ref{fig:4}, and the effect qualitatively survives in
the one-loop result. If, however, $d_1^L$ is already very small,
then $d_0^T$ going to zero has only a small effect on $T_c$,
as is the case in Fig.\ \ref{fig:1}. Which of
these two cases is realized depends on the parameter values.
We finally mention that the first order nature of the magnetic 
transition\cite{us_1st_order,Saxena_etal} adds another mechanism for 
suppressing $T_c$ in the PM phase:
For a sufficiently strong first order transition,
and if the case shown in Fig.\ \ref{fig:4} is realized, then the
effective $t$ may be large enough everywhere for the
system to miss the maximum of $T_c$ in the PM phase, but not in the FM phase.

We gratefully acknowledge discussions with Maria Teresa Mercaldo,
Mario Cuoco, and Rastko Sknepnek. Part of this work was performed
at the Aspen Center for Physics. DB thanks the members of the 
Physics Department at 
Oxford University for their hospitality. This work was
supported in part by the NSF, grant Nos. DMR--98--70597 and
99--75259, by the DFG, grant No. Vo659/3, and by the
EPSRC, grant No. GR/M 04426.

\vfill\eject
\end{document}